\documentclass[prl,showpacs,twocolumn,aps]{revtex4}
\usepackage{graphicx}
\newcommand{\be}{\begin{equation}}
\newcommand{\ee}{\end{equation}}
\newcommand{\bea}{\begin{eqnarray}}
\newcommand{\eea}{\end{eqnarray}}
\newcommand{\up}{\uparrow}
\newcommand{\dn}{\downarrow}
\newcommand{\bwt}{\begin{widetext}}
\newcommand{\ewt}{\end{widetext}}
\begin{document}
\title{
Quantum Mechanics of Spin Transfer in Ferromagnetic Multilayers
}
\author{Wonkee Kim and F. Marsiglio}
\affiliation{
Department of Physics, University of Alberta, Edmonton, Alberta,
Canada, T6G~2J1}
\begin{abstract}
We use a quantum mechanical treatment of a ballistic spin current to describe
novel aspects of spin transfer to a ferromagnetic multilayer.
We demonstrate quantum phenomena from
spin transmission resonance (STR) to
magnetoelectric spin echo (MESE), depending on
the coupling between the magnetic moments in the ferromagnetic thin films.
Our calculation reveals new channels through which the zero spin transfer
occurs in multilayers: the STR and MESE.
We also illustrate that counter-intuitively, a negative spin torque
can act initially on the second moment in a bilayer system.
\end{abstract}

\date{\today}
\maketitle

The phenomenon of spin transfer \cite{slonczewski,berger} 
has resulted in a recent wave of progress
\cite{myers,katine,tsoi,waintal,wegrowe,albert,stiles,zhang,kiselev} 
to study interactions between spin-polarized
electrons and a magnetic moment in a ferromagnetic film. 
However, the physics of spin transfer still has 
not been fully appreciated quantum mechanically. 
The underlying principle for this phenomenon is angular momentum 
conservation. Since the incoming electrons are spin-filtered
by the magnetic moment, as a reaction to the filtering effect
the magnetic moment becomes tilted to align along the direction
of the incoming spin. This alignment is also known as the spin torque
because of the associated moment change. However,
spin is a quantum mechanical concept while torque is a 
classical quantity. These disparate views are reconciled since
spin is transferred quantum mechanically from the incoming electrons to the
magnetic moment, but the effect on the moment appears in a
classical manner.
In this paper, we demonstrate some quantum mechanical aspects of the
spin transfer in a ferromagnetic multilayer by adopting an adiabatic
approximation to describe the dual electron spin/ferromagnetic moment
system.

Two very interesting quantum mechanical phenomena in these problems are
the spin transmission resonance (STR) \cite{str} and 
the magnetoelectric spin echo (MESE) effect \cite{mese}. 
The STR is due to quantum interference
of the right-moving (transmitted) and left-moving (reflected)
electron wave functions in the ferromagnetic film.
On the other hand, the MESE
is a consequence of the time reversal symmetry
between the two magnetic moments.
The STR and MESE have some similarities 
as well as differences between them. 
Both phenomena occur with {\it effectively} zero spin transfer
to the moments. However, as we elucidate below, the physics governing the
two phenomena is quite different.

First, STR
can occur not only in a multilayer system but also in a single 
ferromagnetic film while the MESE can only
take place in a multilayer system. For STR, resonance
of the electron wave function is essential, while for the MESE 
we require
an anti-symmetric configuration between the two magnetic moments
{\it all the time}. In this way the spin current, once 
absorbed by the first
moment, can be generated by the second moment. That is, the second
moment plays the role of a spin battery.
Consequently, a strong interaction between the two moments is necessary
for the MESE while STR is more or less insensitive to the
moment-moment interaction. Moreover
specific values of the kinetic energy of the incoming electrons are required
for STR; this is not the case for the MESE.
   
To account for both STR and the MESE, we consider
a simple model Hamiltonian for two single-domain ferromagnetic films
with all key interactions included
\be
H=\frac{p^{2}}{2m}-2J_{H}\sum_{i=1}^{2}{\bf M}_{i}\cdot{\bf s}
-\frac{J_{M}}{\gamma_{0}}{\bf M}_{1}\cdot{\bf M}_{2}\;,
\label{H}
\ee
where $m$ is the electron mass, $J_{H}$ is the coupling between
the magnetic moment and the incoming electron spins, $J_{M}$ is the coupling
between the moments, and $\gamma_{0}$ is the gyromagnetic ratio.
We consider the ballistic regime as in Refs.~\cite{str,mese}.
%

The magnetic moments ${\bf M}_{1}$ and ${\bf M}_{2}$ in the magnetic 
multilayers are assumed to
originate from the local spins in the same ferromagnets
with a constant magnitude $M_{0}$.
The directions of ${\bf M}_{1}$ and 
${\bf M}_{2}$ at a given time are $(\theta_{1},\;\phi_{1})$ and 
$(\theta_{2},\;\phi_{2})$,
respectively, and are subject to the interactions.
In the Hamiltonian Eq.~(\ref{H}), the second term 
transfers the electron spin to the magnetic moments via the spin flip process
while the third term controls the relative orientation of
${\bf M}_{1}$ and ${\bf M}_{2}$. 

Origins of the coupling between the magnetic moments are
the exchange interaction and a magnetic dipole interaction.
The effective coupling is determined by the competition between the
two interactions depending on the distance between the magnetic moments.
A simple evaluation of the magnetic interaction energy shows 
that the anti-parallel configuration is more stable than
the parallel case when the distance between the two ferromagnetic films
is larger than the atomic scale.
For example, for one-dimensional uniform ferromagnets
of length $L$ separated by the distance $d$, 
the magnetic dipole interaction energy 
is approximately $\pm 2M^{2}_{0}/L^{2}d$
for parallel $(+)$ and anti-parallel $(-)$
configurations.
For two-dimensional uniform ferromagnets with size of $L\times L$,
the energy is about $\pm 8M^{2}_{0}\ln(L/d)/L^{3}$.
As we mentioned earlier, a necessary condition for the
MESE is that the two moments remain anti-parallel {\it for all times}.
Thus 
$J_{M}$ must be negative with magnitude large compared to $J_H$.
Otherwise, the spin torque
will align easily both moments parallel to the initial direction of 
the incoming spin; the MESE can no longer occur.

We show in Fig.~1 the geometry of the problem. 
The incident direction of the electrons is in the positive X
direction, and the two films are placed perpendicular to this direction,
in the YZ plane. Translational symmetry is assumed in the plane.
Therefore, the problem is effectively one-dimensional as in 
Refs.~\cite{stiles,str}; this is
a reasonable approximation in the ballistic regime. 
It is also assumed that 
regions I, III, and V are non-ferromagnetic and are the same in nature
for simplicity. When we solve for the dynamics of the magnetic moments
we use the adiabatic approximation. While this is applicable in most
cases, it should be used with caution for the MESE
because the magnetic moments will oscillate rapidly if the moment-moment
coupling is large. The applicability of the adiabatic approximation
requires the time scale for the electrons to
be much shorter than that of the moments.

\begin{figure}[tp]
\begin{center}
\includegraphics[height=2.6in,width=3.0in]{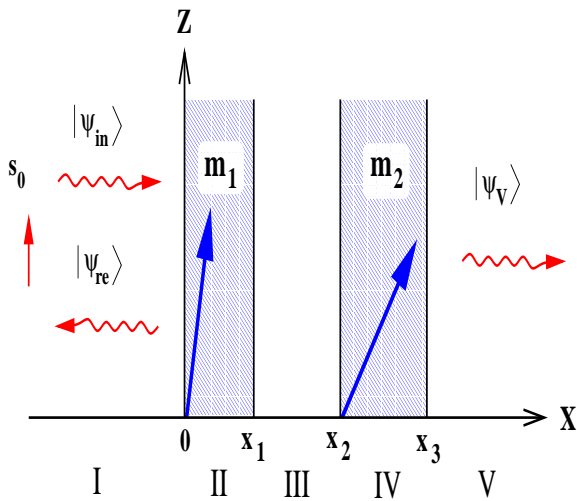}
\caption{(Color online) Quantum mechanical problem associated with
the spin transfer to a ferromagnetic multilayer. 
Here ${\bf m}_{i}={\bf M}_{i}/M_{0}$.
Regions II and IV represent the ferromagnetic films
with thickness $L_{1}$ and $L_{2}$, respectively. 
Note that $x_{1}=L_{1}$, $x_{2}=L_{1}+d$, and $x_{3}=L_{1}+d+L_{2}$
where $d$ is the thickness of a spacer.
Regions I, III, and V are non-ferromagnetic.
}
\end{center}
\end{figure}

In principle, one can use any basis set to represent the wave functions
in each region. For example, one could use the spin-up/down state
$|\pm\rangle$ in the lab frame for the basis.
However, using the {\it right} basis makes the calculations much easier.
We introduce eigenstates $|\chi_{\sigma}\rangle$ and
$|\xi_{\sigma}\rangle$ of the interactions 
$2J_{H}{\bf s}\cdot{\bf M}_{1}$
and $2J_{H}{\bf s}\cdot{\bf M}_{2}$ such that
$2J_{H}{\bf s}\cdot{\bf M}_{1}|\chi_{\sigma}\rangle=
\pm J_{H}M_{0}|\chi_{\sigma}\rangle$ and
$2J_{H}{\bf s}\cdot{\bf M}_{2}|\xi_{\sigma}\rangle=
\pm J_{H}M_{0}|\xi_{\sigma}\rangle$.
Using the eigenstates of the interaction, we can represent
the wave function in each region. We emphasize that 
this procedure can be systematically generalized to
a system with more ferromagnetic films.

The wave functions in regions I, II,III, IV, and V
can be written as follows:
\bea
|\psi_{I}\rangle&=&|+\rangle e^{ikx}
+\sum_{\sigma=\up,\dn}
R_{\sigma}|\chi_{\sigma}\rangle\langle\chi_{\sigma}|+\rangle
e^{-ikx}
\nonumber\\
|\psi_{II}\rangle&=&\sum_{\sigma=\up,\dn}
\left(A_{\sigma}e^{ik_{\sigma}x}+B_{\sigma}e^{-ik_{\sigma}x}\right)
|\chi_{\sigma}\rangle\langle\chi_{\sigma}|+\rangle
\nonumber\\
|\psi_{III}\rangle&=&\sum_{\sigma=\up,\dn}
T_{\sigma}|\chi_{\sigma}\rangle\langle\chi_{\sigma}|+\rangle e^{ikx}
\nonumber\\
&+&\sum_{\sigma=\up,\dn}
R'_{\sigma}|\xi_{\sigma}\rangle\langle\xi_{\sigma}|+\rangle
e^{-ikx}
\nonumber\\
|\psi_{IV}\rangle&=&\sum_{\sigma=\up,\dn}
\left(A'_{\sigma}e^{ik_{\sigma}x}+B'_{\sigma}e^{-ik_{\sigma}x}\right)
|\xi_{\sigma}\rangle\langle\xi_{\sigma}|+\rangle
\nonumber\\
|\psi_{V}\rangle&=&\sum_{\sigma=\up,\dn}
T'_{\sigma}|\xi_{\sigma}\rangle\langle\xi_{\sigma}|+\rangle
e^{ikx}
\eea
The coefficients $R_{\sigma}$ to $T'_{\sigma}$ are determined by the
boundary conditions at $x=0$, $L_{1}$, $L_{1}+d$, and $L_{1}+d+L_{2}$.
In general, the coefficients depend on the directions of the
moments in the magnetic multilayer. If different bases are
used to expand the wave functions of each region, the coefficients
will be changed. However, the wave functions remain the same.
Since expressions of the coefficients
are too long, we do not show them here. Instead, we explain important
properties associated with the coefficients and the wave functions,
below, for the extreme cases: i) the STR with no dipole interaction
and ii) the MESE with a strong dipole interaction.

Under the STR condition, the ferromagnetic film becomes transparent
to the incoming spins \cite{str}. 
One set of conditions that is required is that
$L_1 = 2nL_{00}$ $(n=1,2,\cdots)$ 
for an energy ratio $\eta=5/4$. 
Here $L_{00}=\pi/\sqrt{2mJ_{H}M_{0}}$
is a typical length scale of the ferromagnetic film and
$\eta$ is the ratio of the incoming electron energy $k^{2}/2m$
to the interaction energy $J_{H}M_{0}$. 
If we have the STR condition only for the first film in region II, 
$T_{\up}=T_{\dn}$ so that the spin state of the forward-moving
wave in region III
is the same as that of the incident wave. 
But, unlike the single layer case, 
$R_{\sigma}\ne0$ because the wave function in region III is partially
reflected at $x_{2}=L_{1}+d$ and the reflected wave can then
pass through the first film in reverse (the same STR condition applies).
We point out that even in this case $R_{\sigma}\ne R'_{\sigma}$
because the magnetic moments are not necessarily parallel to each another.
Suppose STR takes place only in
region IV where the second film resides. Then clearly
$R'_{\sigma}=0$ so that no wave is reflected at $x_{2}$.
Nevertheless, $T'_{\up}\ne T'_{\dn}$ because the forward-moving wave in
region III
has both $|+\rangle$ and $|-\rangle$ components.
Each spin component passes through the second film freely 
due to the STR condition in region IV.
This means that the film is transparent to two differently 
spin-polarized electron beams
under the same STR condition. 
When we have the STR condition for both films,
$T_{\up}=T_{\dn}$ as well as $T'_{\up}=T'_{\dn}$ and
$R_{\sigma}=R'_{\sigma}=0$ as expected.

The other interesting limit results in the MESE. In this case
the coupling between moments is very large, and antiferromagnetic.
We find
$T'_{\up}=T'_{\dn}$ so that the net spin transfer is zero
because the spin state in region V is the same as that of the
incoming electrons. It is in 
this sense that the MESE is similar to the STR for both films.
However, if the two films also satisfy the STR condition, then
the moments no longer rotate in response to the spin current,
so the result is more properly thought of as an STR, occurring twice,
once in each film. In all cases we have discussed so far
the flux is conserved
in each region while the spin current is not because
the spin torque acting on the magnetic moment in region II(IV)
is equivalent to the spin current difference between region I
and III (and between III and V). The net spin torque acting on the
multilayer system is the spin current difference between region I
and V. Later, we will describe the dynamics of the magnetic moments
in term of the spin current. 

The equations of motion of the two magnetic moments 
${\bf M}_{1}$ and ${\bf M}_{2}$ are
\bea
\frac{d{\bf M}_{1}}{dt}&=&2J_{H}{\bf M}_{1}\times\langle\psi_{II}|
{\bf s}|\psi_{II}\rangle+J_{M}{\bf M}_{1}\times{\bf M}_{2}
\nonumber\\
\frac{d{\bf M}_{2}}{dt}&=&2J_{H}{\bf M}_{2}\times\langle\psi_{IV}|
{\bf s}|\psi_{IV}\rangle+J_{M}{\bf M}_{2}\times{\bf M}_{1}\;.
\eea
Note that different wave functions are used to evaluate 
the spin expectation values for the equations of the moments.
We use dimensionless units as in Ref. \cite{str}, where
${\bf m}_{i}={\bf M}_{i}/M_{0}$ and the time is $\tau=j_{0}t$,
where $j_{0}=N_{e}L_{00}/(mS_{local})$ is the one-dimensional current
with the number of incoming electrons $N_{e}$ per unit length and
$S_{local}=M_{0}/\gamma_{0}$. In these units, the coupling constant
due to the magnetic dipole interaction becomes 
$\alpha=(S_{local}/2N_{e}L_{00})J_{M}/J_{H}$.
Now the equations of motion for 
${\bf m}_{i}$
become
\bea
\frac{d m_{ix}}{d \tau}&=&-{\beta}_{i}m_{iz}m_{ix}+{\gamma}_{i}m_{iy}
+\alpha\left({\bf m}_{i}\times{\bf m}_{j}\right)_{x}
\nonumber\\
\frac{d m_{iy}}{d \tau}&=&-{\beta}_{i}m_{iz}m_{iy}-{\gamma}_{i}m_{ix}
+\alpha\left({\bf m}_{i}\times{\bf m}_{j}\right)_{y}
\nonumber\\
\frac{d m_{iz}}{d \tau}&=&{\beta}_{i}\left(1-m^{2}_{iz}\right)
+\alpha\left({\bf m}_{i}\times{\bf m}_{j}\right)_{z}\;,
\label{eom}
\eea
where $i,j=1,2$ $(i\ne j)$,
\bea
\beta_{i}&=&
\frac{1}{2}\int{}dx\;\mbox{Im}\left[C^{*}_{i\dn}C_{i\up}\right]
\nonumber\\
\gamma_{i}&=&
\frac{1}{2}\int{}dx\;\mbox{Re}\left[C^{*}_{i\dn}C_{i\up}\right]
\nonumber
\eea
with $C_{1\sigma}=A_{\sigma}e^{ik_{\sigma}x}+B_{\sigma}e^{-ik_{\sigma}x}$,
$C_{2\sigma}=A'_{\sigma}e^{ik_{\sigma}x}+B'_{\sigma}e^{-ik_{\sigma}x}$.
The integration range for $\beta_{i}$ and $\gamma_{i}$ is given by
the thickness of the corresponding ferromagnetic film. 

\begin{figure}[tp]
\begin{center}
\includegraphics[height=2.6in,width=3.0in]{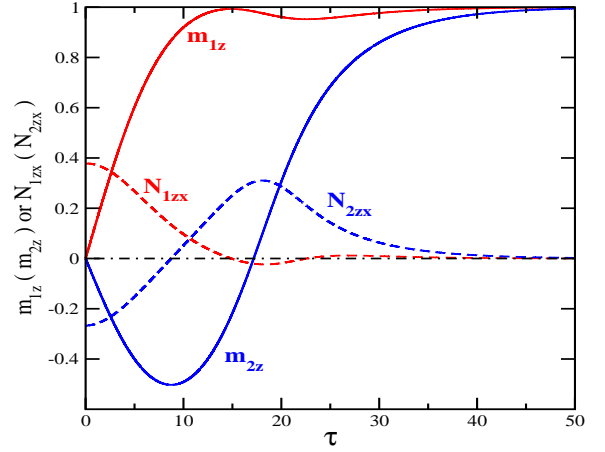}
\caption{(Color online)
The time evolution of the magnetic moments (solid curves) and the
spin torque (dashed curves) on the moments for
$\eta=5/4$, $L_{1}=2.6L_{00}$,
$L_{2}=2.4L_{00}$, and $d=L_{00}$.
No magnetic dipole interaction is included $(\alpha=0)$.
The initial value of the moments is
${\bf m}_{1}=-{\bf m}_{2}=(0,1,0)$.
The spin torque on $m_{1z}$ is
$N_{1zx}=Q_{I,zx}-Q_{III,zx}$ and the torque on
$m_{2z}$ is $N_{2zx}=Q_{III,zx}-Q_{V,zx}$
}
\end{center}
\end{figure}

As we mentioned earlier, another way to understand the dynamics
is using the spin current tensor;
\be
Q_{ij}=\frac{s}{2im}\Biggl[\langle\psi|\sigma_{i}\partial_{j}|\psi\rangle
-h.c\Biggr]\;,
\ee
where $s=1/2$ is the electron spin.
The spin current in a particular region can be calculated using the
wave function in the region. The time evolution of the magnetic moment
is governed by the spin torque which is the net spin flux absorbed
by the moment \cite{stiles}. For example, the time evolution of 
$m_{1z}$ and $m_{2z}$ is determined by $N_{1zx}$ and $N_{2zx}$, respectively,
where $N_{1zx}=Q_{I,zx}-Q_{III,zx}$ and $N_{2zx}=Q_{III,zx}-Q_{V,zx}$.
For a wave function
$|\psi\rangle=
\left(\begin{array}{c}
a_{1}\\
a_{2}
\end{array}\right)e^{ikx}+
\left(\begin{array}{c}
b_{1}\\
b_{2}
\end{array}\right)e^{-ikx}
$,
the spin current becomes
\bea
Q_{xx}&=&\frac{k}{m}\left\{
\mbox{Re}\left[a^{*}_{1}a_{2}\right]-
\mbox{Re}\left[b^{*}_{1}b_{2}\right]\right\}
\nonumber\\
Q_{yx}&=&\frac{k}{m}\left\{
\mbox{Im}\left[a^{*}_{1}a_{2}\right]-
\mbox{Im}\left[b^{*}_{1}b_{2}\right]\right\}
\nonumber\\
Q_{zx}&=&\frac{k}{2m}\left\{\left(|a_{1}|^{2}-|a_{2}|^{2}\right)-
\left(|b_{1}|^{2}-|b_{2}|^{2}\right)\right\}\;.
\eea
The components $a_{i}$ and $b_{i}$ are given by the wave functions
in the regions I, III, and V; namely, $a_{i}$ ($b_{i}$) is the 
forward-moving (backward-moving) component, respectively.
For example, in region III, 
$a_{1}=\sum_{\sigma}T_{\sigma}|\langle+|\chi_{\sigma}\rangle|^{2}$, 
$a_{2}=\sum_{\sigma}T_{\sigma}
\langle-|\chi_{\sigma}\rangle\langle\chi_{\sigma}|+\rangle$, 
$b_{1}=\sum_{\sigma}R'_{\sigma}|\langle+|\xi_{\sigma}\rangle|^{2}$, and
$b_{2}=\sum_{\sigma}R'_{\sigma}
\langle-|\xi_{\sigma}\rangle\langle\xi_{\sigma}|+\rangle$.

We solved Eq.~(\ref{eom}) for a typical case with
$\eta=5/4$, $L_{1}/L_{00}=2.6$, $L_{2}/L_{00}=2.4$, and $d/L_{00}=1$.
The magnetic dipole interaction is set to zero $(\alpha=0)$ but
a small interaction $(|\alpha|<1)$ gives a similar result. The initial
direction of $m_{1}=(0,1,0)$ and $m_{2}=(0,-1,0)$. In Fig.~2, we plot
$m_{1z}$ and $m_{2z}$ as the solid curves with the corresponding labels.
Based on the single layer analogy, one could expect that
${\bf m}_{1}$ and ${\bf m}_{2}$ end up aligning with the electron
spin current along the Z direction.
The behavior of ${\bf m}_{1}$ does follow this naive expectation
except for a dip near $\tau=22$, However, the initial
dynamics of ${\bf m}_{2}$ is unexpected; the magnetic 
moment initially acquires a
moment in the negative Z direction, anti-parallel to the incoming
spin current.

We also plot the spin torque acting on the moments as dashed curves
with labels. The behavior of the torques 
explains
the dynamics of the magnetic moments.
Mostly a positive spin torque is applied to $m_{1}$                
so that
it aligns along the Z direction more or less monotonically 
(there is a small dip in the torque 
which corresponds to the dip in the $m_{1z}$ curve mentioned above).
On the other
hand ${\bf m}_{2}$ moves below the XY plane due to
a {\it negative} torque up to $\tau=8$. Then a {\it positive} torque
begins acting on ${\bf m}_{2}$ until the magnetic moment aligns
to the Z direction.

A second case is one in which a strong magnetic dipole interaction
exists between the two films: MESE.
In Fig.~3, we plot the numerical results
for $\alpha=-500$, $\eta=15$, $L_{1}/L_{00}=L_{2}/L_{00}=2$, and 
$d/L_{00}=1$. We use a large energy for the incoming electron
to ensure the applicability of the adiabatic approximation.
The solid oscillatory curves with labels are the time evolution of the 
magnetic moments. The dashed line is the spin current in region I and
V while the dashed (oscillatory) curve is the spin current in region III.
For the MESE, the spin torque is not completely equivalent to
the spin current difference because the magnetic dipole interaction
also contributes to the torque. Nevertheless, the spin current still
provides useful information for the dynamics of the moments.
As shown in Fig.~3, $Q_{I,zx}-Q_{III,zx}$ is positive.
This indicates that the spin current is absorbed by ${\bf m}_{1}$.
On the other hand, $Q_{III,zx}-Q_{V,zx}$ is negative but
$|Q_{III,zx}-Q_{V,zx}|=|Q_{I,zx}-Q_{III,zx}|$. Consequently,
the exactly same value of the spin current is generated by 
${\bf m}_{2}$ because of the time reversal symmetry between the two moments.
This is why effectively zero spin is transferred for the MESE.

\begin{figure}[tp]
\begin{center}
\includegraphics[height=2.6in,width=3.0in]{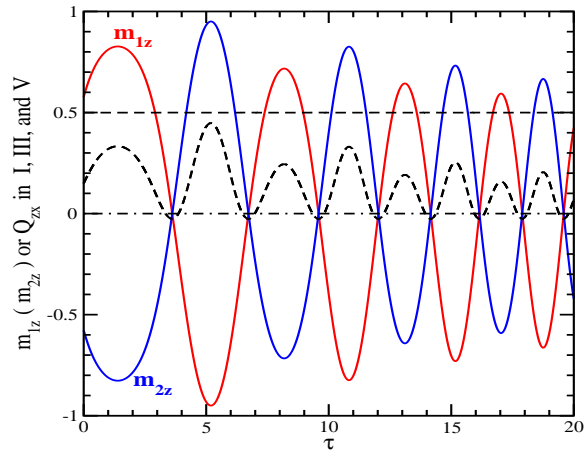}
\caption{(Color online)
The time evolution of $m_{1z}$ and m$_{2z}$ (solid curves)
for MESE with $\alpha=-500$, $\eta=15$, $L_{1}=2L_{00}$,
$L_{2}=2L_{00}$, and $d=L_{00}$.
The dashed line is the spin current in region I and V while
the dashed curve is the spin current in region III.
The initial value of the moments is
${\bf m}_{1}=-{\bf m}_{2}=(1/\sqrt{3},1/\sqrt{3},1/\sqrt{3})$.
}
\end{center}
\end{figure}

As discussed in Ref.~\cite{mese}, the thickness of the two films
should be the same. Brataas {\it et al.} estimated the effect of
the thickness mismatch. We quantify this
effect by calculating $(Q_{I,zx}-Q_{V,zx})/Q_{I,zx}$ for given parameters.
For the same parameters except for $\eta=10$,
we found that a $2\%$ mismatch $(L_{2}/L_{00}=2.04)$ gives $0.3\%$ 
of the spin transfer on the time scale of Fig.~3 $(\tau=20)$
while a $5\%$ mismatch yields $1.7\%$ of the spin transfer on this time scale.
A longer time scale of course leads to more spin transfer. We also found that
the spin transfer is insensitive to $\alpha$ (for large values), and that
if $\eta$ is increased, the spin transfer is decreased for the same time
scale.

In summary we have studied quantum mechanical aspects of the spin transfer
to ferromagnetic bilayers. Our formulation is readily generalized to
multilayers. The physics of the spin transfer in the multilayers
is generically different from that of a single layer.
We demonstrated novel quantum mechanical features 
which can occur only in the multilayer system such as negative
spin torque and magnetoelectric spin echo (MESE). The spin transmission 
resonance (STR) in a multilayer is also illustrated. 
Our calculation reveals new channels through which the zero spin transfer
occurs in multilayers: the STR and MESE.
In spite of irregularities in real materials,
we expect to see some of the qualitative features described here. 

This work was supported in part by the Natural Sciences and Engineering
Research Council of Canada (NSERC), by ICORE (Alberta), and by the
Canadian Institute for Advanced Research (CIAR).
\bibliographystyle{prl}

\begin{thebibliography}{1}

\bibitem{slonczewski} J.C. Slonczewski, J. Magn. Magn. {\bf 159},
L1 (1996); {\bf 195}, L261 (1999)

\bibitem{berger} L. Berger, \prb {\bf 54}, 9353 (1996).

\bibitem{myers} E.B. Myers, D.C. Ralph, J.A. Katine,
R.N. Louie, and R.A. Buhrman, Science {\bf 285}, 867 (1999).

\bibitem{katine} J.A. Katine, F.J. Albert, R.A. Buhrman,
E.B. Myers, and D.C. Ralph, \prl {\bf 84}, 3149 (2000).

\bibitem{tsoi} M. Tsoi, A.G.M. Jansen, J. Bass, W.-C. Chiang,
V. Tsoi, and P. Wyder, Nature {\bf 406}, 46 (2000).

\bibitem{waintal} X. Waintal, E.B. Myers, P.W. Brouwer, and
D.C. Ralph, \prb {\bf 62}, 12317 (2000).

\bibitem{wegrowe} J.-E. Wegrowe, D. Kelly, T. Truong, Ph. Guittienne
and J.-Ph. Ansermet, Europhys. Lett. {\bf 56}, 748 (2001).

\bibitem{albert} F.J. Albert, N.C. Emley, E.B. Myers, D.C. Ralph, and
R.A. Buhrman, \prl {\bf 89}, 226802 (2002).

\bibitem{stiles} M. D. Stiles and A. Zangwill, \prb {\bf 66}, 014407 (2002).

\bibitem{zhang} S. Zhang, P.M. Levy, and A. Fert, \prl {\bf 88}, 236601 (2002).

\bibitem{kiselev} S.I. Kiselev, J.C. Sankey, I.N. Krivorotov, N.C. Emley,
R. J. Schoelkopf, R.A. Buhrman, and D.C. Ralph,
Nature {\bf425}, 380 (2003).

\bibitem{str} W. Kim and F. Marsiglio, \prb {\bf 69}, 1 (2004).

\bibitem{mese} A. Brataas, G. Zarand, Y. Tserkovnyak, and G.E.W. Bauer,
\prl {\bf 91}, 166601 (2003).

\end{thebibliography}

\end{document}